\documentclass{aa}  
\usepackage{graphicx}
\usepackage{longtable,lscape}
\usepackage{lscape}
\usepackage{txfonts}
\begin{document}
\title{ A catalog of planetary nebula candidates in the Sculptor spiral galaxy NGC\,300 
\thanks{Based on observations collected at the European Southern Observatory, VLT, Paranal, Chile, program ID 077.B-0430}
}
\author{Miriam Pe\~na\inst{1}, Jonnathan Reyes-P\'erez\inst{1}, Liliana Hern\'andez-Mart\'inez\inst{2} \and Miguel P\'erez-Guill\'en\inst{1} 
}
\offprints{M. Pe\~na}
\institute{(1) Instituto de Astronom{\'\i}a, Universidad Nacional Aut\'onoma de M\'exico, Apdo. Postal 70264, M\'ex. D. F., 04510 M\'exico\\
(2) Instituto Nacional de Astrof{\'\i}sica, Optica y Electr\'onica, Tonantzinta, Pue., M\'exico\\
\email{miriam@astro.unam.mx,jperez@astro.unam.mx,lilihe@inaoe.mx,jguillen@astro.unam.mx}
 }
%
\titlerunning{ PNe in NGC\,300}
\authorrunning{ Pe\~na et al.}


\abstract
{}
{[\ion{O}{iii}]\,5007 \AA~ on-band  off-band images, obtained with the Very Large Telescope (VLT) and FORS\,2 spectrograph in two zones (center and outskirts) of the spiral galaxy NGC\,300, are analyzed searching for emission line objects. In particular we search for planetary nebula (PN) candidates to analyze their distribution and luminosity properties, to perform follow-up spectroscopy, and  to study  the planetary nebula luminosity function, PNLF.}
{In  the continuum-subtracted  images, a large number of emission line objects were detected. From this sample we selected as PN candidates those objects with stellar appearance and no detectable central star.   [\ion{O}{iii}]\,5007 instrumental magnitudes  were measured and calibrated by using  spectrophotometric data from the follow-up spectroscopy.}
{We have identified more than a hundred PN candidates and a number of compact HII regions. The PN sample is the largest one reported for this galaxy so far. For all the objects  we present coordinates, instrumental [\ion{O}{iii}]\,5007 magnitudes and apparent nebular [\ion{O}{iii}]\,5007  fluxes and magnitudes.  The  [\ion{O}{iii}]\,5007 observed luminosity function for PNe (PNLF) was calculated for the whole sample and for the central and outskirts samples.  The three PNLF are similar within uncertainties.  We  fit the empirical PNLF  to the observed  PNLF for all the samples.  From our best fit for the whole sample we derived a maximum value for the apparent magnitudes of   $m^\star_{5007}=22.019 \pm 0.022$ and we obtained a tentative estimate of the distance modulus
 m$_{5007}-$M$_{5007}$\,=\,26.29$^{+0.12}_{-0.22}$ mag, which agrees  well with the recent value derived  from Cepheid stars.}
{}
\maketitle
\section{Introduction}

Planetary nebulae (PNe) constitute the evolutionary end point  of low-intermediate mass stars  with initial masses in the range from 1 to 8 M$_\odot$. Their central stars  are post-AGB objects whose UV radiation  ionizes the nebula ejected during the AGB stage. Due to  the fact that PNe emit selectively in a small number of strong and narrow emission lines, they can be discovered at significant distances within the nearby Universe (at least up to 30 Mpc). The  study of PNe provides accurate information on the luminosity, age, metallicity, and dynamics of the parent stellar population and this makes them very useful to test a number of theories about the evolution of  low-intermediate mass stars and their influence in galaxies.

PNe in external galaxies are also useful as distance indicators through the  [\ion{O}{iii}]\,5007 \AA~planetary nebulae luminosity function  (PNLF). Ciardullo et al. (1987) and  Jacoby  (1989) reported that the bright end of the PNLF seems invariant  for all the galaxies they investigated, thus they proposed that the PNLF can be used as distance indicator  if  a complete PN sample in the 2 or 3 brightest magnitudes is obtained. The advantage of this method is that we can see bright PNe in galaxies of all kind of Hubble types, and PNe are easily identified. Although more recently there is some   evidence that the standard empirical  PNLF does not fit well in some  irregular galaxies, like the cases of the SMC presented by Jacoby \& De Marco (2002), the LMC discussed by Reid  \& Parker (2010) and NGC\,6822, analyzed by Hern\'andez-Mart{\'\i}nez \& Pe\~na (2009), the bright end of the PNLF continues to be a robust secondary distance indicator in all kind of galaxies (for spiral galaxies see the work by Herrmann et al. 2008).

 The Scd spiral galaxy NGC\,300 is located in the Sculptor Group at a distance of 1.88 Mpc (distance modulus of 26.37$\pm$0.08) as derived from Cepheid stars  by Gieren et al. (2005). Due to its proximity to the Milky Way,  this galaxy has been the subject of numerous studies related to its stellar content  in the central and outskirt regions. The abundance patterns of its massive stars and HII regions, as well as  the radial abundance  gradients, have been analyzed recently (Urbaneja et al. 2005; Bresolin et al. 2009 and references therein). Surveys for PNe have been reported by Soffner et al. (1996) and Rizzi et al. (2006), but only Soffner et al. published a list with coordinates and characteristics of 34 PN candidates. 
 
 The main aim of this work is to present the PN candidates detected with  [\ion{O}{iii}]\,5007 \AA~ on-band off-band technique, during a deep survey looking for PN candidates in two zones (central and outskirts) of NGC\,300 and to investigate the PN population in this galaxy. We also  built and analyzed  the PNLF.
Images were obtained as a "pre-imaging'' program for  follow-up spectroscopy  of a sample of emission line objects.  Here we present the results of the imaging and part of the spectral data which is used to validate the identification of objects as PNe or HII regions and to calibrate our instrumental [\ion{O}{iii}]\,5007 magnitudes. Complete analysis  of the spectroscopy will be presented in a second paper where line  fluxes, physical conditions and   chemical compositions of observed objects will be analyzed (Pe\~na et al., in preparation).
     
  The paper is organized   as follows:  in \S 2 we present the observations and data reduction.  In \S 3, the criteria to separate PN candidates from other emitting objects and the obtained samples are described. Also comparative photometry and flux calibration for the detected objects  are carried out in this section. In \S 4 we discuss the observed   PNLF constructed from the [\ion{O}{iii}]\,5007 apparent  magnitudes and the fit of the empirical PNLF to the observed ones.  Our results are summarized in \S 5.

  \section{Observations and data reduction}
  
\subsection{Imaging}

[\ion{O}{iii}]  5007 \AA~on-band off-band imaging was obtained in service mode, at the Very Large Telescope, VLT, UT1 (Antu) of the European Southern Observatory (ESO), Paranal, Chile, equipped with the spectrograph FORS\,2 (Appenzeller et al., 1998), on 2006-07-04. This constituted the pre-imaging of program ID 077.B-0430. The filters FILT\_500\_5+85 (on-band filter, central wavelength 5000 \AA, FWHM 50 \AA) and  OIII/6000+52 (off-band filter, central wavelength 5105 \AA, FWHM 61 \AA) were used. 

Two zones were observed. The first one, centered at RA= 00$^{\rm h}$ 54$^{\rm m}$ 49.00$^{\rm s}$, Dec= $-37^o$41$'$02.0$''$, corresponds to the center of the galaxy and the second zone, with central position RA= 00$^{\rm h}$ 55$^{\rm m}$ 22.00$^{\rm s}$, Dec= $-37^o$43$'$00.0$''$, is in the outskirts. For each zone, two on-band images of  375 s  exposure time each, and two off-band exposures of 260 s each, were acquired.  The covered area in each zone  is 6.8$\times$6.8 arcmin$^2$ (corresponding to the field of view of FORS\,2), with a spatial scale of {0.25$''$/pix (standard resolution)}. In Table 1 we list the individual images obtained which, for each position, were slightly dithered for a better coverture of the gap between both CCDs of FORS\,2.

\begin{table}[t]
\centering  
\caption{On-band off-band images obtained for NGC\,300 with FORS\,2}             
\begin{tabular}{l c c}        
\hline\hline                 
\multicolumn{3}{l}{Central zone, center  RA=00$^{\rm h}$ 54$^{\rm m}$ 49.0
$^{\rm s}$, Dec= $-37^o$41$'$02.0$''$}\\
\hline 
frame & filter &  ET$^1$  \\    
\hline                        
r.FORS2.2006-07-04T07 40 50.519\&520 & FILT$\_$500$\_$5 & 375 \\      
r.FORS2.2006-07-04T07 47 45.890\&891 & FILT$\_$500$\_$5& 375 \\
r.FORS2.2006-07-04T07 54 51.223\&224 & OIII/6000 & 260  \\
r.FORS2.2006-07-04T07 59 52.257\&258 & OIII/6000& 260 \\
\hline \hline
\multicolumn{3}{l}{Outskirt zone, center  RA=00$^{\rm h}$ 55$^{\rm m}$ 22.0$^{\rm s}$, Dec= $-37^o$43$'$00.0$''$}\\
\hline 
r.FORS2.2006-07-04T08 09 36.304 \&305& FILT$\_$500$\_$5 & 375 \\
r.FORS2.2006-07-04T08 17 11.560 \& 561 & FILT$\_$500$\_$5 & 375 \\
r.FORS2.2006-07-04T08 24 55.767 \& 768& OIII/6000 & 260 \\
r.FORS2.2006-07-04T08 29 58.301 \& 302 &  OIII/6000 & 260 \\
\hline                                   
\multicolumn{3}{l}{$^1$ Exposure time in seconds}
\end{tabular}
\end{table}

Images were reduced and calibrated with the normal procedures of the ESO pipeline. That is, reduced images are bias subtracted, flat-fielded, and astrometrically calibrated, but they are not photometrically calibrated. According to FORS1+2 User«s Manual (VLT-MAN-ESO-13100-1543, Issue 3) the astrometric precision is  one pixel, equivalent to 0.25 arcsec. During the observations the seeing conditions were better than 0.9 arcsec and the sky transparency was clear.

 Each set of reduced  images of Table 1 was  first aligned and then recombined using IRAF{\footnote{IRAF is distributed by the National Optical Astronomy Observatories, which is operated by the Association of Universities for Research in Astronomy, Inc., under contract to the National Science Foundation.} routines. The resulting images are  equivalent to a 12.5 min exposure for the [\ion{O}{iii}] on-band filter  and a 8.7 min exposure for the off-band filter. In Fig. 1, we present a combined image showing the central and outskirt  [\ion{O}{iii}] on-band images. There is a small overlap between both zones. The analysis of the images is presented in \S3.

\begin{figure*}[ht] 
\begin{center}
\label{NGC300-image}
\includegraphics[width=14cm,height=9cm]{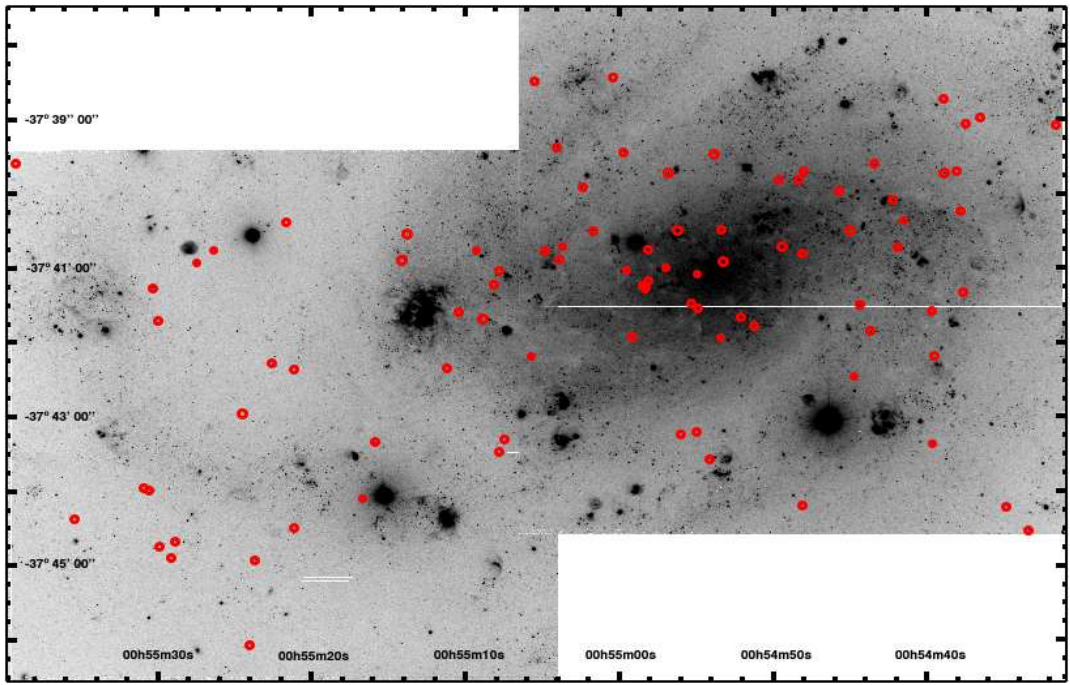}
\caption{Images in [\ion{O}{iii}]$\lambda$5007 observed for the center and outskirt zones, put together.  R.A. and Dec coordinates are marked for J2000. North is up. The detected PN candidates are marked.}
\end{center}
\end{figure*}

\subsection{Spectroscopy and calibration of imaging}

 Follow-up spectroscopy  for more than 40 objects (that were classified as PN candidates and compact HII regions in our imaging),  was carried out with the same telescope and instrument (VLT and FORS2 in MXU mode, program ID 077.B-0430(B)) on 2006 August 19 - 22. Although a complete analysis of the spectrophotometric data will be presented elsewhere, here we describe the observations and present partial results   because the spectroscopic data helps to confirm the  nature of the analyzed objects (PN or compact HII region) and we  use the calibrated [O{\sc iii}] 5007 fluxes obtained spectroscopically, to calibrate our imaging data (see Hern\'andez-Mart{\'\i}nez et al. 2009 for a complete description of this procedure).

 The grisms 600B, 600RI and 300I were used to cover a spectral range from about 3600 to 9500 \AA.  Several frames were obtained  for each grism; total exposure times were of about 3.3 to 3.5 h for spectra with grisms 600B and 600RI and about 1.5 h for spectra with grism 300I. The slit width was 1$''$ for all the objects and the spectral resolution varied from about 0.7 \AA ~ to 1.2 \AA.
Data were reduced and calibrated  using the ESO pipeline and IRAF routines. The standard stars EG274, LDS749B and BMP16274  were observed each night, through a slit of 5$''$ width, for flux calibration. During the spectroscopic run the sky was clear and the seeing conditions varied from 0.7$''$ to 0.9$''$, however some flux could have been lost  in the 1$''$ slit. The losses can amount  to 10-15\% which should be considered in the uncertainties  of our flux calibration. 

\section{Analysis and calibration of the images}
 To search for emission line objects, the  on-band and off-band combined images were scaled (by scaling some stars present in both images), and then they were subtracted. This allows  to reduce the effect of crowding and the emission line objects are more easily detected in the subtracted image. In addition we used  the  ``blinking''  technique of the on-band and off-band images, which is very powerful to detect, by eye, faint emitting objects. The continuum-subtracted [\ion{O}{iii}]\,5007 images show numerous extended and compact emitting objects, from which we selected all the possible  PN candidates and some compact HII regions. The criteria to distinguish between both type of objects are described   in the next section. 

\subsection{Separating PN candidates from other  emission line objects}

 PN candidates were selected following the prescriptions given in Pe\~na et al. (2007) and references therein. That is, PN candidates should be point-like objects (at the distance of NGC\,300, a 1 pc diameter  nebula  would appear as a 0.1$''$ size object, thus, not resolved) and the central star should not be detected. This is not  the case of central stars of HII regions, because for a distance modulus of  26.37 mag,  an O9.5\,V$-$B0\,V star (the faintest able of producing a very low excitation HII region), possessing an absolute magnitude $M_{\rm V}  \sim -3.90$ (Martins et al.  2005) has an apparent magnitude V$\sim$ 22.5 mag, which translates into m(5007) $\sim$ 25.6 mag for the narrow-band filter we used (actually the observed magnitude would be  of about 26 mag, when the reddening to NGC\,300 is considered).  Our faintest [\ion{O}{iii}]\,5007 magnitude detected is about 27.7 mag, therefore central stars of HII regions can be detected in our images, but  PN central stars are typically 2$-$2.5 mag fainter (M\'endez et al. 1992), thus they are undetectable.  However it could happen that a field star is projected on a nebula, complicating  the classification. Only spectroscopy can help in such a situation (see e.g., the cases discussed by Richer \& McCall 2007).

Certainly there are intruders in any sample of PN candidates. This could be the case of very compact HII regions with a faint central star or no central star at all (as it could occur in compact knots  embedded in  HII regions). Also very young unresolved supernova remnants (SNR) can masquerade as PNe, however these objects are rare, therefore their contamination is minimal. In addition, objects from the background like unresolved  Ly$\alpha$ galaxies  with extremely strong Ly$\alpha$ could mimic PNe. At z=3.126, Ly$\alpha$ is redshifted to 5009 \AA~which is  the redshifted wavelength of [\ion{O}{iii}]\,5007 for NGC\,300  by considering a heliocentric radial velocity of  142 km s$^{-1}$ (de Vaucouleurs et al. 1991). However the surface density of such objects is low. In a deep search  in an area of 0.27 deg$^2$, Gronwald et al. (2007) detected 162 Ly$\alpha$ emitters with monochromatic fluxes, in the 5007 \AA ~band, brighter than 1.5$\times$10$^{-17}$ erg cm$^{-2}$ s$^{-1}$. Of these, about 60 objects  would be above our detection limit  which is $\sim$ 2.5$\times$10$^{-17}$ erg cm$^{-2}$ s$^{-1}$. As we sampled a field of 6.8$\times$ 6.8 arcmin$^2$, only 3 of such  objects  would be contaminating  our sample. Thus,  these contaminants would represent only a few \% of our sample,  and the majority would be  in the faint end of the [\ion{O}{iii}]\,5007 flux distribution. 

 For the case of  compact HII regions (cHII), ideally a second or even a third discriminant criterion would help.
For instance, the excitation degree represented by the [\ion{O}{iii}]\,5007\,/\,H$\alpha$ flux ratio, which is usually larger in PNe compared to HII regions, is a commonly used criterion to select PN candidates.  Magrini et al. (2000), Ciardullo et al. (2002), and Herrmann et al. (2008) have selected as PNe those objects with [\ion{O}{iii}]\,5007\,/\,H$\alpha \geq$ 3.
Another criterium could be to consider that cHII should be much brighter in H$\alpha$ than  PNe, due to the larger amount of ionizing photons emitted by  their central stars.  Unfortunately we only have [\ion{O}{iii}]\,5007 images, and this latter criterion  is not true for this  emission line, as it depends strongly on the nebular  ionization structure, and cHIIs could be bright or faint in 5007 \AA.  Then, 
we  only depend on the absence of a visible  central star  and the compactness of objects to separate PN candidates from cHII, therefore we have been very careful in selecting as PNe those strictly stellar objects (size as PSF) with no detected central star. Possibly some very compact knots embedded in HII regions  could be contaminating the sample. These knots  have in general higher density than the surrounding nebula, thus they  show low excitation and low [\ion{O}{iii}] 5007, therefore it would be difficult to detect them in an [\ion{O}{iii}] 5007 image unless they are very bright. Due to this we consider  that the contamination of our PN sample with this kind of knots  represents no more than a few \%.

\subsection{Results from imaging}
 In conclusion, from our deep search and following the criteria given above, we detected more than a hundred PN candidates which is the largest sample obtained for NGC\,300 so far. Their coordinates are presented in columns 2 and 3 of Table 2 and their spatial distribution is shown in Fig. 1. Interestingly, PN candidates appear concentrated mainly in the central zones of the galaxy but they are not particularly associated with the spiral structure where, on the other hand, most of the HII regions reside. There are large zones in the outer regions of the galaxy where no PN has been found. This is mainly due to  the low stellar density  in these zones. 
 
 We also detected a large number of extended and compact HII regions; the latter appear in our imaging as compact nebulae with a detectable central star. 
 A brief sample of cHII were chosen to be analyzed spectroscopically together with a sample of PN candidates (see \S2.2). Some of our cHII have been previously reported in the catalogue by Deharveng et al. (1988).
 The physical  conditions and abundances of these regions will be discussed in
a following paper  in comparison with the spectroscopic properties of the PN candidates.

  We rediscovered 25 of the 34 PN candidates reported by Soffner et al. (1996) who analyzed some central zones of NGC\,300, all of which lie inside our central zone. The 25 rediscovered objects correspond to the brightest PN candidates of Soffner et al., numbered in their work  from 1 to 23 and their objects \#25 and \#27. In this work, we confirmed the PN nature of all of them except their object \#4, which from our spectroscopy resulted to be  a compact HII region (it is our object \#39 of Table 2), their objects \#16 and \#20 which appear as  faint and diffuse nebulae (therefore   they are not PNe) and their object \#27 which is a faint star. The latter three are not included in our list. For the rest of Soffner et al. objects,  they  definitely do not appear in our images even when we searched carefully in the positions of the alleged PNe.
 Our [\ion{O}{iii}] recombined images are deeper than the ones by Soffner et al.  because ours have 12.5 min of exposure time with the VLT which is equivalent to more than one hour  with a 3.6-m telescope which were the exposure time and telescope used by Soffner et al.  Besides  our images were obtained under much better seeing conditions (better than 0.9 $''$ against the 1$''$ up 2$''$ reported by Soffner et al.) and they have better spatial resolution, thus we were able to detect much fainter objects than object \#25 of Soffner et al., therefore we conclude that their  objects \#s 24, 26, and from \#28 to \#34 are possible   misidentifications.  It is worth to notice that the missing objects would belong to their field W, which is very crowded and it was observed with only half an hour of exposure time, therefore the misidentifications are not rare.} Soffner et al. (1996) ID's are included in column 9 of our Table 2.

 For all the PN candidates (104 objects) and 8 compact HII regions  we have performed  a comparative photometry measuring the instrumental magnitudes in the    [\ion{O}{iii}]\,5007 on-band image. The IRAF task   {\it digiphot.apphot.phot} was used. The typical PSF of the images is 2.9 pix (FWHM), thus an aperture of 5 pix radius (equivalent to 0.63$''$) was used to integrate the object flux and  the sky was subtracted from a ring of 5 pix radius and width of 2 pix around the object.  The instrumental magnitudes m$_i$(5007) and their errors are presented in columns 4 and 5 of Table 2.

\subsection{[\ion{O}{iii}]   calibration} 

The most  confident  way to distinguish among  different types of emission nebulae  is by their spectral analysis. However this is expensive in terms of telescope time because in external galaxies these objects are very faint. Our follow-up spectroscopy included more than 40 nebulae. From these data we confirm the PN nature of 19 PN candidates of the central zone and 13 of the outskirts. The confirmed PNe are those  objects having a measured [\ion{O}{iii}]\,5007 flux in column 6 of Table 2. Several  compact HII regions were also observed. Their calibrated [\ion{O}{iii}]\,5007 \AA ~ fluxes are also listed in column 6  of Table 2. 

Fig. 2 presents the relation between our instrumental magnitudes m$_i$(5007) and the logarithm of spectroscopic fluxes for all the PN candidates and compact HII  regions analyzed. It is notable  that a very good linear correlation can be fitted to the data, through 5 magnitudes.  The two brightest objects lying slightly below the correlation are the cHIIs \#5 and \#87 of Table 2. These cHIIs are slightly extended and some  flux can have been lost in the spectroscopic slit of 1$''$ width. On the other extreme, the four faintest  objects which do not fit very well are very faint PNe for which the instrumental magnitudes have relatively large errors ($\Delta$m $\geq$ 0.2 mag). To discard these objects  produces no significant difference in the correlation.   

The linear fit shown in Fig. 2   was  used to calculate calibrated [\ion{O}{iii}]\,5007 fluxes for all the objects from their instrumental magnitudes. The fluxes are presented in column 7 of Table 2 and in column 8 we list the [\ion{O}{iii}]\,5007 apparent magnitudes, calculated as m(5007) = $-$2.5 log F$_{5007}  -$13.74, with F$_{5007}$ in erg cm$^{-2}$ s$^{-1}$ (Allen 1973). 

\begin{figure}[ht] 
\begin{center}
\label{magnitudes}
\includegraphics[width=7cm]{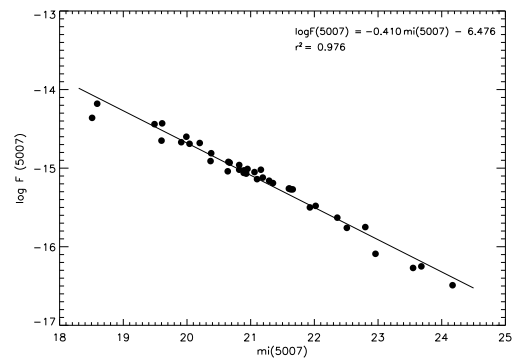}
\caption{ Relation between spectroscopic [\ion{O}{iii}]\,5007 fluxes, F(5007), and instrumental magnitudes m$_{\rm i}$(5007)   for all the spectroscopically analyzed objects (PN candidates and compact HII regions are included). The parameters of the linear fit are shown in the figure. The   objects lying  slightly out of the correlation, are discussed in the text.}\label{fig2}
\end{center}
\end{figure}

\section{The PN sample and the planetary nebulae luminosity function }

There are 66 PN candidates in the central zone and 34 in the outskirts, for which  confident [\ion{O}{iii}]\,5007 apparent magnitudes were determined. Although  the number of objects is  small  for a reliable statistics in order to derive the peak magnitude of the PNLF and  the distance to NGC\,300, a rough [\ion{O}{iii}]\,5007 luminosity function (PNLF) can be constructed for  the whole sample and for both zones. The PNLF helps to obtain a better understanding of the PN populations  in a galaxy; thus, with this purpose, in the following sections we construct and analyze  the differential PNLF.

\begin{figure*}[ht] 
\begin{center}
\label{PNLF}
\includegraphics[width=18cm,height=7cm]{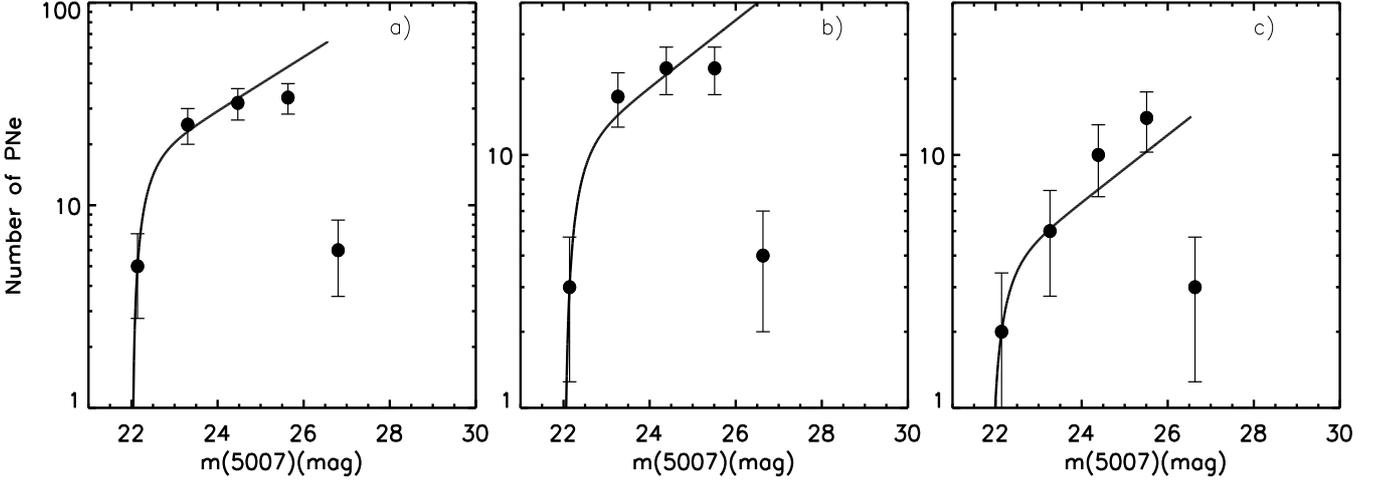}
\caption{Observed PN luminosity function for: a) the whole sample, b) the PNe in the central zone, and (c): the PNe in
 the outskirts of NGC\,300.  Bin sizes are described in the text.  The errors bars correspond to the
Poissonian  error. Solid lines represent our best fit of the empirical PNLF for the  4 brightest bins in each case.}

\end{center}
\end{figure*}

\subsection{The PNLF behavior}

In our experience, the shape of the  PNLF is very sensitive to the selection of the bin size,  in particular when the sample is small.  With a too small bin size, some bins could have very few objects and   a very sparse PNLF could be obtained. The effect of the bin size can be observed, for instance, in the two PNLFs  built for the SMC PNe by Jacoby et al. (1998) and Jacoby \& De Marco (2002); the first one, with a bin size of 0.2 mag, is very sparse and the other, with a bin of 1 mag, shows a shape similar to the empirical PNLF.  A full discussion of the effect of the bin size on the PNLF will published elsewhere (Rodr{\'\i}guez-Gonz\'alez et al. in preparation).

Thus, due to the low number of objects in our samples we were cautious with the bin size selection and carried out several proofs for the bin size (defined as  $bs=(m_{\mathrm{max}}-m_{\mathrm{min}})/Nbins$, where {\it bs} is the bin size, $m_{\mathrm{max}}$ and $m_{\mathrm{min}}$ are the maximum and minimum magnitude values in the sample, and {\it Nbins} is the number of bins that we assign to our distribution),  in order to obtain an observed PNLF with a shape similar to the empirical one (eq. 2). Finally, from our [\ion{O}{iii}]\,5007 calibrated magnitudes for the PN sample  we computed the observed  PNLFs for the whole sample  and for each region (inner and outskirts) using  bin sizes of 1.16 mag, 1.12 mag and 1.12 mag respectively. Therefore, for all the cases, the sample is divided in 5 bins. 
The three observational PNLFs are presented in Fig. 3.  The error bar assigned to each bin represents the statistical Poissonian error.

 The fit of the empirical PNLF  to the observed PNLF  (computed by using a fitting scheme based in the Levenberg-Marquardt technique which uses  a $\chi^2$ minimization) was calculated by converting the usual differential luminosity function for absolute magnitudes given in eq. 1 (Jacoby 1989; Ciardullo et al. 1989),
 
 \begin{equation}
N(M) \propto {\rm e}^{0.307M_{5007}} (1- {\rm e}^{3(M ^*_{5007}- M_{5007})})   
 \end{equation}

 \noindent to a luminosity function in apparent magnitudes (eq. 2)  
 \begin{equation}
N(m) = N {\rm e}^{- 0.307\mu} {\rm e}^{0.307m_{5007}} (1- {\rm e}^{3(m^*_{5007} - m_{5007})}) , 
 \end{equation}

\noindent where  $N$ is a normalization constant, $\mu$ is the apparent distance modulus, $\mu = 5\, {\rm log\, d - 5 + A_{5007}}$, and $M^*_{5007}$ and $m^*_{5007}$ are the absolute and apparent peak magnitudes of the luminosity function.

 The observed PNLF for the whole sample (Fig. 3 left) seems complete in the first 4 bins (the fifth one already shows incompleteness)  thus we used them 
 for fitting the empirical PNLF (eq. 2).  The fit is shown as a solid line in the figure. A Kolmogorov-Smirnov test  applied  to this fit  shows that both functions (observed and empirical) are equal within the first 4 bins, with a reliance level  of 99.96\%. The PNLF adjusted for the whole sample returns a value m$^*$(5007)= 22.019$\pm$0.022 mag, and a value N e$^{-0.307\mu}$ = 0.0187$\pm$0.0021.

 For the central zone, the observed PNLF (Fig. 3 middle) also seems complete in the first 4 bins and this PNLF is very similar to the one for the whole sample because it contains the majority of the total PN candidates.  The fitting of the empirical PNLF  is shown as a solid line. Again a Kolmogorov-Smirnov test  applied  to this fit  indicates that the observed and empirical functions are equal within the first 4 bins, with a reliance level  of 99.96\%. The PNLF adjusted for the central sample returns values of m$^*$(5007)= 22.02$\pm$0.03 mag, and  N e$^{-0.307\mu}$ =0.0118 $\pm$ 0.00158, completely similar to values of the whole sample.
   
 The sample of the outskirts (Fig. 3 right)  has only 34 objects and the empirical PNLF was adjusted to  the first 4 bins which seem complete. Also a Kolmogorov-Smirnov test for this fit predicts that both curves are the same, with a reliance of 99.96\%. The m$^*$(5007) value returned is 21.88$\pm$0.07  mag, and  N e$^{-0.307\mu}$ = 0.0042 $\pm$ 0.00105, which is slightly lower than the previous values, but it could be consider similar within uncertainties.
 
 Therefore we find no evidence of a significant  change in the PNLF with the galactic position in NGC\,300.

 The behavior of the PNLF is normal, in the sense that no one presents any noticeable dip as the ones found in the irregular galaxies, e.g., the SMC (Jacoby \& De Marco 2005), the LMC (Reid \& Parker 2010) and NGC\,6822 (Hern\'andez-Mart{\'\i}nez et al. 2009), which have been interpreted as due to the existence of a younger population of PNe in which the central star evolution proceeds very quickly.  
\smallskip 

As the top 3 magnitudes of the PNLF of the  whole sample are significantly complete and this is the distance-sensitive segment, it is tempting to  use our fit result for a tentative  estimate of the distance to NGC\,300. As already said,  the apparent peak magnitude returned by our best fit   is m$^*$(5007) = 22.019$\pm$0.022.
This value should be dereddened for the foreground extinction.  It is known that the external reddening towards NGC\,300 is very low. The estimated values fluctuate between E(B-V)= 0.013 (Schlegel et al. 1998) and E(B-V) = 0.096 (Gieren et al. 2005). These values  translate in an absorption A$_{\rm 5007}$ of  0.05 and 0.3 mag respectively. By adopting A$_{\rm 5007} \sim$\,0.20 and assuming a peak absolute magnitude M$^{\star}$(5007)= $-$4.47 as derived from  Fig. 5 of Ciardullo et al. (2002) for the NGC\,300 central metallicity (adopted to be 12+log O/H = 8.57, Bresolin et al. 2009), we obtain a distance modulus m$-$M=26.29$^{+0.12}_{-0.22}$. The error bar includes the formal error of the fit (0.022 mag),  the error in A$_{\rm 5007}$ (about 0.1 mag) and, on the bright side, we have added a 10\%  error due to possible flux losses in  the spectroscopic slit. Our derived distance   is  in   good agreement, with the distance modulus of 26.37$\pm$0.08 reported from Cepheids by Gieren et al. (2005). This confirms  the robustness of the method of deriving distances to external galaxies through the bright cut-off of the PNLF.

\section{ Conclusions}

\hskip 0.6cm 1) From deep [\ion{O}{iii}]\,5007  on-band and off-band imaging, performed with VLT FORS\,2 to search  for line emission  objects in two zones of the spiral galaxy NGC\,300, we detected more than a hundred PN candidates, and many compact and extended HII regions. From our follow-up spectroscopy, the 32 candidates analyzed, of the 104 objects in the sample, have been confirmed as true planetary nebulae, therefore we are confident that our methods for selecting PNe are appropriate.

 2) The instrumental [\ion{O}{iii}]\,5007  magnitudes derived from the imaging were calibrated by using results from our spectroscopy. This allowed us to determine calibrated apparent  [\ion{O}{iii}]\,5007  magnitudes for our PN candidates and some compact HII regions.

  3) We built the observed  [\ion{O}{iii}]\,5007 PNLF for the whole PN sample as well as for the samples of the central zone and the outskirts. All the PNLFs are similar within uncertainties and they appear normal, in the sense that they show no dip as the ones found in the LMC, the SMC and NGC\,6822. We fit the empirical  PNLF to our data, obtaining a very good fit for the 3 brightest magnitudes in all the cases,  in particular the whole sample and  the central one return the same results.  
   
 4) Our best fit for the 3 brightest magnitudes of the observed PNLF for the whole sample allows us to determine a distance modulus  m$-$M=26.29$^{+0.12}_{-0.22}$ mag which  is in  good agreement with the value of  26.37$\pm$0.08 mag derived by Gieren et al. (2005) for Cepheid stars in NGC\,300.

\begin{acknowledgements}
Invaluable comments  by G. Stasi\'nska  are deeply appreciated.  M. Pe\~na is grateful to DAS, Universidad de Chile, for hospitality during a sabbatical stay when part of this work was performed.  L. H.-M. benefited from the hospitality of the Departamento de Astronom{\'\i}a, Universidad de Chile.  L. H.-M., J. R.P., and M. P.-G.  received a scholarship from  CONACYT-M\'exico. 
 M. P. gratefully acknowledges financial support from FONDAP-Chile and DGAPA-UNAM. This work received financial support from grants \#43121 (CONACYT-M\'exico),  IN-112708 and IN-105511(PAPIIT DGAPA-UNAM).
  \end{acknowledgements}

%
\addtocounter{table}{2}
\longtabL{2}{
\begin{longtable}{rccccccll}
\caption{ \hskip 3cm {\bf Table 2} Photometric properties of PN candidates and compact HII regions in NGC 300. \label{tbl-2}}\\
\hline \hline
No.  &  R. A.(2000)   &  Dec(2000)  &  m$_i$(5007)  &  error  & log F$_{5007}^1$  & log F$_{5007}^2$  &m(5007) & other ID$^3$,  \\
 & h ~~~m~~~~s~~~~ & $^o$~~~~$'$~~~~~$''$ & mag & mag & & &mag & comments \\
\endfirsthead
\caption{\hskip 3cm {\bf Table 2.} continued.}\\
\hline \hline
No.  &  R. A.(2000)  &  Dec(2000)  &  m$_i$(5007)  & error &   log F$_{5007}^1$  & log F$_{5007}^2$    & m(5007) & other ID$^3$,   \\
 & h ~~~m~~~~s~~~~ & $^o$~~~~$'$~~~~~$''$ & mag & mag & & &mag&comments \\
\hline
\endhead
\hline
\endfoot
 \hline
1 & 00 54 31.63 & -37 39 04.25 & 23.45 & 0.20 &  & -16.091 & 26.49 &  \\
2 & 00 54 33.17 & -37 38 28.00 & 20.37 & 0.02 & -14.913 & -14.830 & 23.33 & cHII\\ 
3 & 00 54 33.42 & -37 44 31.64 & 21.03 & 0.02 &  & -15.097 & 24.00  \\  
4 & 00 54 34.85 & -37 44 12.16 & 23.38 & 0.10 &  & -16.062 & 26.42  \\ 
5 & 00 54 35.39 & -37 39 36.00 & 18.59 & 0.01 & -14.179 & -14.099 & 21.51 & cHII, D\,37\\  
6 & 00 54 35.62 & -37 41 16.80 & 24.04 & 0.18 &  & -16.334 & 27.09  \\   
7 & 00 54 35.75 & -37 41 10.14 & 23.73 & 0.12 &  & -16.206 & 26.77  \\
8 & 00 54 36.05 & -37 39 50.80 & 19.49 & 0.01 & -14.438 & -14.468 & 22.43 & cHII, D\,38 \\    
9 & 00 54 36.57 & -37 38 58.70 & 24.70 & 0.80 &  & -16.603 & 27.77   \\ 
10 & 00 54 37.55 & -37 39 03.49 & 23.06 & 0.17 &  & -15.931 & 26.09   \\ 
11 & 00 54 37.71 & -37 41 18.98 & 24.42 & 1.41 &  & -16.488 & 27.48   \\ 
12 & 00 54 37.88 & -37 40 14.05 & 21.93 & 0.04 & -15.496 & -15.467 & 24.93   \\ 
13 & 00 54 38.13 & -37 39 41.40 & 23.70 & 0.12 &  & -16.193 & 26.74   \\ 
14 & 00 54 38.91 & -37 39 43.20 & 20.67 & 0.01 & -14.934 & -14.951 & 23.64   \\ 
15 & 00 54 38.94 & -37 38 43.48 & 23.74 & 0.12 &  & -16.209 & 26.78 \\ 
16 & 00 54 39.59 & -37 42 10.69 & 23.18 & 0.15 &  & -15.979 & 26.21 \\ 
17 & 00 54 39.68 & -37 43 21.40 & 22.51 & 0.08 & -15.764 & -15.703 & 25.52\\ 
18 & 00 54 39.76 & -37 41 34.64 & 23.37 & 0.22 &  & -16.058 & 26.40 \\ 
19 & 00 54 40.04 & -37 40 01.99 & 23.93 & 0.18 &  & -16.288 & 26.98 \\ 
20 & 00 54 41.58 & -37 40 21.40 & 22.02 & 0.10 & -15.480 & -15.505 & 25.02 & S\,12 \\ 
21 & 00 54 41.93 & -37 40 43.07 & 22.69 & 0.10 &  & -15.779 & 25.71 & S\,15 \\ 
22 & 00 54 42.23 & -37 40 04.80 & 20.82 & 0.02 & -15.023 & -15.012 & 23.79 & S\,3 \\ 
23 & 00 54 43.47 & -37 39 36.11 & 21.38 & 0.03 &  & -15.242 & 24.36 \\ 
24 & 00 54 43.70 & -37 41 51.29 & 20.93 & 0.02 & -15.069 & -15.056 & 23.90 & S\,8 \\ 
25 & 00 54 44.42 & -37 41 29.40 & 20.82 & 0.02 & -14.955 & -15.012 & 23.79 & S\,2  \\ 
26 & 00 54 44.78 & -37 42 27.72 & 22.67 & 0.08 &  & -15.771 & 25.69 \\ 
27 & 00 54 45.03 & -37 40 28.82 & 22.99 & 0.12 &  & -15.902 & 26.01 & S\,17 \\ 
28 & 00 54 45.77 & -37 41 30.40 & indef  &  &  &  &  & S 22 \\ 
29 & 00 54 45.78 & -37 39 58.11 & 22.47 & 0.10 &  & -15.687 & 25.48 \\ 
30 & 00 54 45.97 & -37 37 52.84 & 24.06 & 0.37 &  & -16.341 & 27.11\\
31 & 00 54 48.05 & -37 39 41.67 & 23.29 & 0.20 &  & -16.026 & 26.33 \\ 
32 & 00 54 48.12 & -37 40 48.57 & 24.01 & 0.37 &  & -16.321 & 27.06 & S\,23 \\ 
33 & 00 54 48.19 & -37 44 11.51 & 22.20 & 0.04 &  & -15.578 & 25.21 \\ 
34 & 00 54 48.20 & -37 43 41.76 & 25.74 & 0.93 &  & -17.029 & 28.83\\ 
35 & 00 54 48.38 & -37 39 48.42 & 20.04 & 0.01 & -14.693 & -14.692 & 22.99 \\ 
36 & 00 54 49.23 & -37 40 19.14 & 25.08:  & 1.00 &  & -16.757 & 28.15:  \\ 
37 & 00 54 49.44 & -37 40 42.49 & 22.86 & 0.12 &  & -15.849 & 25.88 \\ 
38 & 00 54 49.71 & -37 39 48.43 & 22.49 & 0.10 &  & -15.696 & 25.50 \\ 
39 & 00 54 51.25 & -37 41 46.21 & 20.64 & 0.02 & -15.039 & -14.939 & 23.61 & S\,4, cHII in HII \\ 
40 & 00 54 52.08 & -37 42 43.20 & 20.89 & 0.02 & -15.056 & -15.042 & 23.86 \\ 
41 & 00 54 52.14 & -37 41 39.91 & 22.45 & 0.13 &  & -15.681 & 25.46 \\ 
42 & 00 54 53.26 & -37 40 54.41 & 22.59 & 0.10 &  & -15.737 & 25.60 & S\,14 \\ 
43 & 00 54 53.40 & -37 41 56.15 & 22.96 & 0.10 &  & -15.889 & 25.98 \\ 
44 & 00 54 53.42 & -37 40 28.92 & 21.90 & 0.05 &  & -15.455 & 24.90 \\ 
45 & 00 54 53.82 & -37 39 27.50 & 21.10 & 0.02 & -15.144 & -15.127 & 24.08 \\ 
46 & 00 54 54.15 & -37 43 33.92 & 22.28 & 0.08 &  & -15.610 & 25.29 \\ 
47 & 00 54 54.32 & -37 42 32.36 & 23.61 & 0.25 &  & -16.156 & 26.65 \\ 
48 & 00 54 54.91 & -37 41 32.42 & 20.91 & 0.20 & -15.025 & -15.049 & 23.88 & S\,7\\ 
49 & 00 54 54.97 & -37 41 04.56 & 23.09 & 0.20 &  & -15.943 & 26.12 \\ 
50 & 00 54 54.98 & -37 43 11.64 & 24.45 & 0.40 &  & -16.498 & 27.51 \\ 
51 & 00 54 55.33 & -37 41 28.54 & 20.89 & 0.02 & -15.025 & -15.041 & 23.86 & S\,5 \\ 
52 & 00 54 55.99 & -37 43 14.70 & 23.50 & 0.25 &  & -16.111 & 26.54 \\ 
53 & 00 54 56.26 & -37 40 29.64 & 20.17 & 0.01 &  & -14.746 & 23.12 \\ 
54 & 00 54 56.83 & -37 39 43.49 & 21.19 & 0.02 & -15.118 & -15.164 & 24.17 \\ 
55 & 00 54 56.97 & -37 40 59.77 & 23.02 & 0.26 &  & -15.914 & 26.05 \\ 
56 & 00 54 57.22 & -37 41 22.92 & 24.75 & 0.64 &  & -16.624 & 27.82 \\ 
57 & 00 54 57.42 & -37 41 00.96 & 19.61 &  0.01& -14.431 & -14.517 & 22.55 & cHII, D\,101 \\ 
58 & 00 54 58.12 & -37 40 44.87 & 20.65 & 0.02 & -14.916 & -14.941 & 23.61 & S\,1 \\ 
59 & 00 54 58.16 & -37 41 10.32 & 21.87 & 0.04 &  & -15.443 & 24.87 & S\,11 \\ 
60 & 00 54 58.34 & -37 41 16.08 & 21.39 & 0.06 &  & -15.247 & 24.38 & S\,13 \\ 
61 & 00 54 58.48 & -37 41 14.39 & 22.28 & 0.07 &  & -15.611 & 25.29 & S\,10 \\ 
62 & 00 54 59.53 & -37 41 01.46 & 22.52 & 0.08 &  & -15.708 & 25.53 & S\,18 \\ 
63 & 00 54 59.72 & -37 39 26.14 & 21.35 & 0.03 & -15.202 & -15.230 & 24.33 \\ 
64 & 00 55 00.47 & -37 38 26.70 & 23.40 & 0.12 &  & -16.070 & 26.44 \\ 
65 & 00 55 01.71 & -37 40 29.39 & 21.61 & 0.04 & -15.259 & -15.336 & 24.60 & S\,9 \\ 
66 & 00 55 02.44 & -37 39 54.65 & 21.06 & 0.02 & -15.052 & -15.111 & 24.04 \\ 
67 & 00 55 03.12 & -37 42 08.10 & 23.49 & 0.15 &  & -16.106 & 26.52 & S\,21 \\ 
68 & 00 55 03.72 & -37 40 42.49 & 23.19 & 0.17 &  & -15.984 & 26.22 & S\,25 \\ 
69 & 00 55 03.97 & -37 40 53.33 & 20.95 & 0.02 & -15.008 & -15.066 & 23.92 & S\,6 \\ 
70 & 00 55 04.04 & -37 42 38.95 & 23.33 & 0.23 &  & -16.041 & 26.36 \\ 
71 & 00 55 04.79 & -37 40 44.70 & 23.70 & 0.15 &  & -16.193 & 26.74 \\ 
72 & 00 55 04.85 & -37 40 46.27 & 23.02 & 0.14 &  & -15.916 & 26.05 & S\,19 \\ 
73& 00 55 05.51 & -37 38 29.04 & 23.53 & 0.16 &  & -16.123 & 26.56 \\ 
74& 00 55 05.77 & -37 42 11.88 & 20.38 & 0.02 & -14.814 & -14.832 & 23.34 &\\  
75 & 00 55 05.81 & -37 42 39.96 & 23.11 & 0.23 &  & -15.949 & 26.13 \\ 
76 & 00 55 07.21 & -37 41 42.29 & 22.82 & 0.10 &  & -15.833 & 25.84 \\ 
77 & 00 55 07.49 & -37 43 18.40 & 23.14 & 0.12 &  & -15.962 & 26.16 \\ 
78 & 00 55 07.76 & -37 41 02.01 & 23.38 & 0.13 &  & -16.062 & 26.42 \\ 
79 & 00 55 07.78 & -37 43 28.30 & indef    \\ 
80 & 00 55 08.11 & -37 41 13.22 & 23.81 & 0.17  &  & -16.236 & 26.85 \\ 
81 & 00 55 08.13 & -37 39 38.66 & 23.93 & 0.25 &  & -16.288 & 26.98   \\ 
82 & 00 55 08.94 & -37 41 40.99 & 22.76 & 0.10 &  & -15.809 & 25.78   \\
83 & 00 55 09.33 & -37 40 10.72 & 23.50 & 0.17 &  & -16.111 & 26.54 \\  
84 & 00 55 09.84 & -37 39 55.42 & 24.07 &  0.34 &  & -16.345 & 27.12   \\ 
85 & 00 55 10.43 & -37 41 34.87 & 23.88 &  0.18&  & -16.266 & 26.93   \\ 
86 & 00 55 11.21 & -37 42 20.95 & 22.96 & 0.12 & -16.088 & -15.890 & 25.99   \\ 
87 & 00 55 13.77 & -37 41 39.23 &  18.51& 0.02 & -14.363 &  -14.065& 21.42  & cHII, D137c \\  
88 & 00 55 13.78 & -37 40 32.63 & 22.36 & 0.10 & -15.627 & -15.644 & 25.37   \\ 
89 & 00 55 14.14 & -37 40 52.88 & 22.89 & 0.02  & & -15.859 & 25.91   \\ 
90 & 00 55 15.08 & -37 44 14.5 &  19.60&  0.02 & -14.648 & -14.513 & 22.54 & cHII  \\ 
91 & 00 55 15.91 & -37 43 20.38 & 21.29 &  0.03& -15.157 & -15.206 & 24.27   \\ 
92 & 00 55 16.65 & -37 44 05.93 & 22.80 & 0.10 & -15.749 & -15.823 & 25.82   \\ 
93 & 00 55 21.17 & -37 42 21.63 & 25.49:  & 0.95 &  & -16.926  & 28.58 :   \\ 
94 & 00 55 21.18 & -37 44 29.69 & 23.68 & 0.17 & -16.252 & -16.186 & 26.73   \\ 
95 & 00 55 21.62 & -37 40 22.44 & 23.55 & 0.11 & -16.271 & -16.133 & 26.59   \\ 
96 & 00 55 22.54 & -37 42 16.51 & 21.60 &  0.03& -15.263 & -15.333 & 24.59   \\ 
97 & 00 55 23.68 & -37 44 55.43 & 21.63 & 0.03 & -15.273 & -15.342 & 24.62   \\ 
98 & 00 55 24.06 & -37 46 04.26 & 24.50 & 0.30 &  & -16.522 & 27.56  \\ 
99 & 00 55 24.49 & -37 42 57.43 & 24.69 &  0.35&  & -16.597 & 27.75   \\ 
100 & 00 55 25.87 & -37 43 53.87 &23.62 & 0.17 & &  -16.160 & 26.66 \\
101 & 00 55 26.39 & -37 40 45.05 & 24.17 & 0.18 & -16.485 & -16.387 & 27.23 \\
102 & 00 55 27.20 & -37 43 43.55 &19.91  & 0.02 & -14.672 & -14.640 & 22.86 & cHII  \\   
103 & 00 55 27.49 & -37 40 55.13 & 19.99 &  0.02& -14.603 & -14.670 & 22.94 \\ 
104 & 00 55 28.88 & -37 44 40.13 & 21.66 &  0.03& -15.273 & -15.355 & 24.65 \\ 
105 & 00 55 29.15 & -37 44 53.48 & 23.06 & 0.16 &  & -15.932 & 26.09 \\ 
106 & 00 55 29.93 & -37 44 44.23 & 23.90 &  0.18&  & -16.273 & 26.94 \\ 
107 & 00 55 29.99 & -37 41 42.18 & 24.02 & 0.20 &  & -16.324 & 27.07 \\
108 & 00 55 30.30 & -37 41 15.68 & 20.20 & 0.03 & -14.679 & -14.759 & 23.16 \\ 
109 & 00 55 30.53 & -37 43 59.16 & 21.16 &  0.03& -15.023 & -15.153 & 24.14 \\ 
110 & 00 55 30.90 & -37 43 57.04 & 22.96 &  0.10 &  & -15.891  & 25.99 \\ 
111 & 00 55 35.42 & -37 44 21.93 & 24.14 & 0.25 &  & -16.373 & 27.19 \\ 
112 & 00 55 39.13 & -37 39 35.24 & 24.28 &  0.25&  & -16.431 & 27.34 \\ 
\hline
\hline
\multicolumn{8}{l}{$^1$ Flux from spectroscopy, in erg cm$^{-2}$ s$^{-1}$.}\\
\multicolumn{8}{l}{$^2$ Flux calculated from the equation in Fig. 2, in erg cm$^{-2}$ s$^{-1}$.}\\
\multicolumn{9}{l}{$^3$ S\#:  IDs for PNe from Soffner et al. (1996). }\\
\multicolumn{9}{l}{~~~D\#:  ID«s for compact HII regions (cHII) from Deharveng et a. (1998).}\\

\end{longtable}

}

\end{document}